\documentclass[12pt]{article}
\usepackage[dvips]{graphicx}

\renewcommand{\L}     {{\cal L}}
\renewcommand{\bar}[1]{\overline{#1}}

\newcommand{\eg}     {{\em e.g.}}
\newcommand{\cf}     {{\em cf.}}
\newcommand{\VEV}[1]{\left\langle{#1}\right\rangle}

\newcommand{\lowsim}[1]{\,\,\mathrel{\rlap{\lower7pt\hbox{$\sim$}}
                    \hskip1pt\hbox{$#1$}}\,}
\newcommand{\Dslash}{\not{\hbox{\kern-4pt $D$}}}

\begin{document}

\begin{flushright}
{\small
SLAC--PUB--9257\\
October 2002\\ }
\end{flushright}

\vfill
\begin{center}
{{\bf\LARGE Cosmology and the Standard Model}\footnote{Work
supported by Department of Energy contract DE--AC03--76SF00515.}}

\bigskip
James D. Bjorken \\
{\it Stanford Linear Accelerator Center \\
Stanford University, Stanford, California 94309 \\
E-mail:  bjorken@slac.stanford.edu}
\medskip
\end{center}

\vfill

\begin{center}
{\bf\large Abstract}
\end{center}

We consider the properties of an ensemble of universes as function
of size, where size is defined in terms of the asymptotic value of
the Hubble constant (or, equivalently, the value of the
cosmological constant). We assume that standard model parameters
depend upon size in a manner that we have previously suggested,
and provide additional motivation for that choice. Given these
assumptions, it follows that universes with different sizes will
have different physical properties, and we estimate, very roughly,
that only if a universe has a size within a factor $\sqrt 2$ of
our own will it support life as we know it. We discuss
implications of this picture for some of the basic problems of
cosmology and particle physics, as well as the difficulties this
point of view creates.

\vfill

 \newpage

\section{Introduction}
\label{sec:1}

Our universe seems, according to the present-day evidence, to be
spatially flat and to possess a nonvanishing cosmological constant
\cite{ref:a}. These features, while not yet rock-solid
experimentally, are hardly what would have been anticipated by the
founding fathers of cosmology. The cosmological constant in
particular is, for cosmologists and general relativists, the Great
Mistake. And for elementary particle physicists it is the Great
Embarrassment. It is fair to say that each school would just as
soon see it go away. But in this paper we assume that it will not
do so, and that the present evidence will prevail.

The cosmological constant is a peculiar quantity. By definition it
has something to do with cosmology. But it also has something to
do with the local structure of elementary particle physics, where
it represents the stress-energy density $\mu$ of the vacuum.
\begin{equation}
\L_{cc} = \int d^4x\, \sqrt{-g}\, \mu^4 = \frac{1}{8\pi G} \int
d^4x\, \sqrt{-g}\, \Lambda \ . \label{eq:a}
\end{equation}
Instead of the parameter $\mu$, the cosmologist will use
$\Lambda$, as defined above, but expressed in terms of the
properties of the spacetime of our distant future, a future
dominated by dark energy and exponential expansion:
\begin{equation}
ds^2 = d\hat t^2 - e^{2H_\infty\hat t} (d\hat r^2 + \hat r^2\,
d\theta^2 + \hat r^2\, \sin^2\theta\, d\phi^2) \ . \label{eq:ab}
\end{equation}
This matter-free spacetime is equivalent to static DeSitter space,
characterized by a horizon radius $R_\infty$,
\begin{equation}
ds^2 = \left(1-\frac{r^2}{R^2_\infty}\right)\, dt^2 -
\left(1-\frac{r^2}{R^2_\infty}\right)^{-1} dr^2 - r^2
(d\theta^2+\sin^2\theta\, d\phi^2)
\label{eq:b}
\end{equation}
via the coordinate transformations
\begin{eqnarray}
r &=& \hat r\, e^{H_\infty \hat t} \nonumber \\
t &=& \hat t - \frac{1}{2H_\infty}\ \ell n (1-H^2_\infty r^2) \ .
\label{eq:bb}
\end{eqnarray}
This horizon radius is determined in terms of the asymptotic Hubble constant
\begin{eqnarray}
H_\infty &=& \lim_{t\rightarrow\infty} H(t) \nonumber \\
R^{-1}_\infty &=& H_\infty = \sqrt{\Omega_\Lambda} \, H_0
\label{eq:c}
\end{eqnarray}
which in turn is determined by gravitational dynamics of the
vacuum energy characterized by $\mu$.
\begin{equation}
\frac{1}{R^2_\infty} = H^2_\infty = \frac{8\pi G}{3}\  \mu^4 =
\frac{\Lambda}{3} \ . \label{eq:d}
\end{equation}

We see from the above equations that, despite the formally
infinite extent of the spatially flat Friedmann-Robertson-Walker
expansion universe, Eq. (\ref{eq:ab}), the presence of a
nonvanishing cosmological constant provides a way of ascribing an
intrinsic, observer-independent, size parameter to our
universe.\footnote{\baselineskip=14pt Given $\Omega=1$, the only
other objective choice of size parameter would seem to be to
utilize one of the several landmark times characterizing the
history of our universe, \eg\ the time of electroweak or strong
phase transition, of matter-radiation equality, or of decoupling.
Our choice is that of matter-dark energy equality, and appears to
us to be the most fundamental.}

\begin{figure}[htb]
\centering
\includegraphics{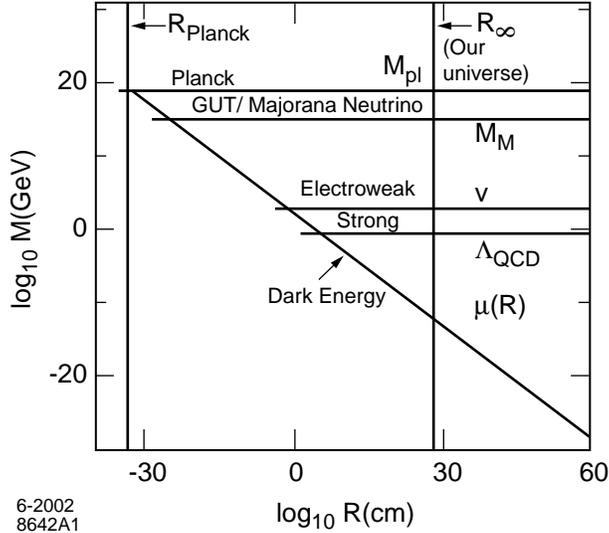}
\caption[*]{\baselineskip=14pt Dependence of fundamental constants
on the horizon size $R_\infty$ of the universe according to
conventional wisdom. \label{fig:1}}
\end{figure}

In this paper we shall be considering an ensemble of universes
similar to our own but with different intrinsic sizes
$R_\infty$.\footnote{\baselineskip=14pt Note that $R_\infty$ is
{\em not} the scale-size $R(t)$ characterizing the expansion of
our own Robertson-Walker universe.  We are {\em not} assuming that
the fundamental constants are time-dependent.  Each universe in
the ensemble undergoes its own Big Bang, and is characterized by
distinct values of standard-model parameters.} The ambivalence
between the elementary particle view of the cosmological constant
as vacuum energy/pressure and the cosmological view of it as a
size parameter for the universe is sharpened by looking at it from
this viewpoint. In particular, conventional wisdom would say that
all of the basic parameters of the standard model, such as
$\Lambda_{QCD}$ or the electroweak vacuum condensate $v$, are to
the best of our knowledge independent of each other. This means
that they are also independent of the vacuum energy/pressure
characterized by the scale $\mu$, since the cosmological term is
just another term in the standard-model Lagrangian density. This
in turn implies that $\Lambda_{QCD}$ and $v$ are also independent
of the size $R_\infty$ of the universe. By definition this is not
the case for the cosmological term ({\em cf.} Fig.~\ref{fig:1}).
For universes smaller than our own, the vacuum energy density
grows. And for universes smaller than about 10 km, the vacuum
energy density exceeds 1 GeV/fm$^3$, the energy scale of the QCD
vacuum.

Does this matter at all? It is at least arguable that it does. To
really understand the vacuum state is one of the most important
goals of fundamental theory, and it is often presumed that one
must go to unified theories, such as string theory, to attain true
enlightenment. But this would imply that, for the vacuum state,
the cosmological degrees of freedom talk to the elementary
particle degrees of freedom such as quark/gluon or Higgs in an
essential way. This point of view is reinforced by the fact that
the dark-energy term in the action is formally renormalized by the
quantum corrections contributed by all the other terms in the
action.

If there is an interconnection between dark energy and QCD vacuum
fluctuations, we might suspect that interesting things occur when
the cosmological vacuum energy scale and the QCD vacuum energy
scale become comparable. Perhaps there is a discontinuous change,
such as occurs for QED at the electroweak scale. Or perhaps the
QCD scale does not wait for such a catastrophe to occur, but
changes continuously as the size parameter of the universe
changes, in a way which is similar to the way the cosmological
constant itself changes.  It is this latter option we entertain in
this paper, an option we have in fact already suggested
\cite{ref:b}. What we assume is, first, that {\em  all
dimensionful parameters $X$ of the standard model may vary with
$R_\infty$, but that to leading approximation they are straight
lines in a log-log plot,  i.e. they satisfy a simple
renormalization-group equation}
\begin{equation}
R_\infty \frac{\partial X}{\partial R_\infty} = -\frac{1}{2}\ \mu
\frac{\partial X}{\partial\mu} = p_X   X + \cdots\label{eq:e}
\end{equation}
We shall discuss the unspecified corrections and other details a
little more in Section \ref{sec:2}.
\bigskip

\begin{figure}[htb]
\centering
\includegraphics{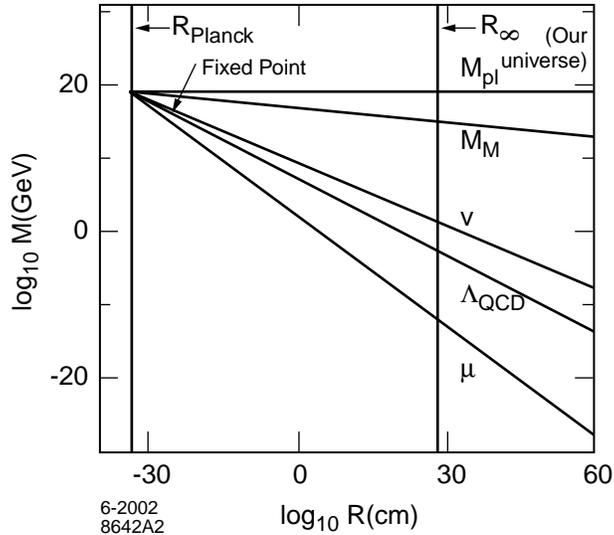}
\caption[*]{\baselineskip=14pt Dependence of the fundamental
constants on the horizon size $R_\infty$ according to the scaling
assumptions. \label{fig:2}}
\end{figure}

By itself the above assumption includes the conventional-wisdom
option, illustrated in Fig.~\ref{fig:1}, and contains little news.
However we in addition assume that {\em  all fundamental
dimensional parameters $X$ flow toward a fixed point occurring for
universes of approximately Planck/GUT size, and that this is the
only such fixed point.} 
Note that only two parameters are needed to describe the gross
dependence of $X$ on $R_\infty$, and that they are fixed by the
value of $X$ at the fixed point and the value of $X$ observed by
us.  This is illustrated in Fig.~\ref{fig:2}.

The first part of the fixed-point assumption should not be any
surprise, since we do expect new physics to occur at the
Planck/GUT scale. The second part of the assumption is really what
provides the motive power for the remainder of this note, and it
is not much more at this stage than an application of Occam's
razor. In detail there are probably exceptions to the rule, but
perhaps insight can be gained even in the absence of being able to
apprehend the exceptions. But leaving aside the possible, even
probable complications, what we have at this point is a
description of the ensemble of universes we are considering,
characterized by the value of $R_\infty$, in terms of modified
standard-model parameters. And we again emphasize that {\em there
are no extra parameters which have been introduced.} Therefore we
can hope to explore in principle the properties of such universes,
using well-defined extrapolations of the laws of chemistry, atomic
physics, nuclear physics, {\em etc.}, with no extra arbitrary
assumptions. In particular we can explore the ``bandwidth" of
features possessed by our own universe. That is, we may try to
determine the minimum and maximum sizes $R_\infty$ for which
nuclear matter exists, or for which hydrogen-burning stars exist,
or for which elements as heavy as carbon exist and are produced.
These and other examples will be discussed in Section \ref{sec:3},
where we shall estimate that if the radius of a universe in our
ensemble of universes is within a factor $\sqrt2$ of our own, the
conditions of life as we know them appear to be satisfied.

Up to this point we may regard the ensemble of universes under
consideration as an abstract set, {\em a la} Gibbs, and the study
of their properties as an abstract intellectual exercise, perhaps
of value in the long run in understanding either microphysics
beyond the standard model, or the macrophysics of our visible
universe. However, there is clear motivation to go further, and to
presume that such an ensemble actually exists. In particular we
may assume that our universe is one member of a multiverse, with
the remaining members causally disconnected from us, as discussed
extensively by Rees \cite{ref:c} and others \cite{ref:d}.

If such a multiverse ensemble really exists, then a primary
quantity of interest is the number distribution $n(R_\infty)$ of
universes of a given size, defined as
\begin{equation}
R_\infty  \frac{dN}{dR_\infty} \equiv n(R_\infty) \ . \label{eq:f}
\end{equation}
As mentioned above, we will roughly determine in Section
\ref{sec:3} the bandwidth $\Delta R_\infty/R_\infty$ within which
the conditions for life as we understand it exist; it is of order
1. Then if
\begin{equation}
\Delta N = n(R_\infty)\ \frac{\Delta R_\infty}{R_\infty} \gg 1
\label{eq:g}
\end{equation}
we may argue that it is not improbable that life should exist in
the multiverse. This is just the condition
\begin{equation} n (R_\infty) \gg 1
\label{eq:gc}
\end{equation}
which appears not to be a heavy constraint. The above line of
argument, and concomitant set of problems, parallels the lines of
argument used to understand our place in our own universe. Why do
we live on Earth rather than Mercury or Pluto? The former is too
hot, the latter too cold. Is our existence improbable, in the
sense that the parameters characterizing Planet Earth are very
finely tuned? The simplest answer is that if life as we know it
exists elsewhere in the universe, {\em i.e.} there is a
sufficiently large population of planets to allow the replication
of conditions found on Earth, no fine tuning is required. The jury
is still out with respect to what that answer is \cite{ref:BW}.
But we may argue that the question is, at least in principle, a
scientific question.  And indeed the hypothesis of a multiverse
softens the above constraint to only require that the multiverse
contain planets with conditions suitable for supporting life as we
know it.

There is an even more specific---and speculative---scenario which
can be entertained and which is discussed in Section \ref{sec:5}.
It is a reductionist version of evolutionary cosmology as
envisaged by Smolin \cite{ref:e}, utilizing a speculative model of
black hole interiors dubbed gravastars by Mazur and Mottola
\cite{ref:f,ref:g}. In this scenario, the interiors of mature
black holes are nonsingular and described by the aforementioned
static DeSitter metric which characterizes the future of our own
universe. This strongly suggests a cosmology of nested black
holes. The interiors of the black holes in our universe comprise
daughter universes, within which there are granddaughter
black-hole universes, {\em etc.} Going in the opposite direction,
we may surmise that our universe consists of the interior of a
black hole existing in a mother universe, which in turn is
embedded within a grandmother universe, {\em etc.} An important
issue in this picture of cosmology is the determination of the
various species of daughter universes (supermassive
galactic-center black holes, stellar-collapse black holes,
$\ldots$) and their size distribution relative to the size of the
parent. Other important parameters are the fertilities of mothers,
{\em i.e.} the number of daughter universes created per mother, as
a function of $R_\infty$ and species. We will try to estimate
these parameters from data and astrophysical theory, and then try
to estimate, for example, the number of sister universes there are
in the multiverse, and thereby to re-examine questions posed
above, such as estimating the number of planets, galaxies, and/or
universes within the multiverse which might support life as we
know it.

By now we have clearly entered a highly speculative level. Indeed
we have organized this note such that at the beginning of each new
section readers making it to that point can become dismissive and
bail out. But in the hope that there is at least one person left
reading this paragraph, we continue on.

Why all this speculation? From the point of view of this writer,
it is motivated by the gravastar scenario, and the related ideas
of emergent gravity and emergent standard model, as advocated by
Volovik \cite{ref:h} and others \cite{ref:i}. The vacuum is
visualized as similar to a quantum liquid such as helium at low
temperatures. In the ``gravastar" scenario, the blackhole
universes are droplets of the quantum liquid, with order
parameters which depend on the size of the droplet. The
cosmological constant is small because in the ground state of the
liquid droplet the pressure (which is measured by the cosmological
constant!) vanishes, up to surface corrections. In the picture
advocated in this note it is not only the cosmological-constant
term in the standard-model Lagrangian density which is a
size-dependent order parameter, but all the others as well. Indeed
for an infinite universe, characterized by an infinite value of
the DeSitter horizon radius $R_\infty$, the entire standard-model
Lagrangian trivializes to a free field theory \cite{ref:b}. All
standard-model interactions are therefore viewed as dependent upon
the existence of a boundary to our universe. Evidently in the
opposite Planck/GUT limit everything becomes strongly coupled.

The crux of this set of ideas lies in the development of a {\em
 microscopic} theory along these lines. And the construction of
such a theory may be aided by having a rough picture of the most
likely cosmological context for these ideas. It is this which is
our primary motivation. But there are a host of obstacles. Some of
these are taken up in Section \ref{sec:6}, which is devoted to
lessons learned and to conclusions, such as they are.

\section{Standard Model Parameters}\label{sec:2}

The fundamental premise of this paper was already stated above Eq.
(\ref{eq:e}), and we expect this hypothesis to be most accurate
for the dimensionful parameters most closely associated with
vacuum energy. These are the cosmological-constant scale $\mu$,
$\Lambda_{QCD}$, the electroweak condensate value $v$, and
probably a large mass scale associated with neutrino mass, in
particular the masses $M$ of the heavy gauge-singlet Majorana
particles associated with the see-saw mechanism of neutrino mass
generation. These masses appear to be in the range
$10^{13}-10^{15}$ GeV, near the GUT scale. For all these
quantities, we assume that Eq. (\ref{eq:e}) holds to good
accuracy.

The fact that $\Lambda_{QCD}$ varies with $R = R_\infty$
(hereafter we drop the subscript) leads to an important
consequence, namely that the strong coupling constant
$\alpha_s(q^2)$ must also vary with $R$. Since
\begin{equation}
\frac{1}{\alpha_s(q^2)} \cong b_s \ \ell n\
\frac{q^2}{\Lambda^2_{QCD}} \qquad b_s = \frac{33-2n_f}{12\pi}
\label{eq:h}
\end{equation}
and
\begin{equation}
\frac{M^2_{pl}}{\Lambda^2_{QCD}} \cong (M^2_{pl} R^2)^{p_s} \qquad
p_s \approx \frac{2}{3} \label{eq:i}
\end{equation}
it follows that
\begin{equation}
\frac{1}{\alpha_s(q^2,R^2)} = b_s \, p_s\, \ell n\, M^2_{pl} R^2 -
b_s\, \ell n\ \frac{M^2_{pl}}{q^2} .
 \label{eq:j}
\end{equation}
What we have is a new renormalization-group equation for
$\alpha_s$
\begin{equation}
R^2 \frac{\partial}{\partial R^2}\ \left(\frac{1}{\alpha_s}\right)
= b_s\, p_s + O(\alpha_s) \label{eq:k}
\end{equation}
which we may compare with the usual expression for the running of
$\alpha_s$ with $q^2$,
\begin{equation}
q^2 \, \frac{\partial}{\partial q^2}\,
\left(\frac{1}{\alpha_s}\right) = b_s + O(\alpha_s) \ .
\label{eq:l}
\end{equation}

\begin{figure}[htb]
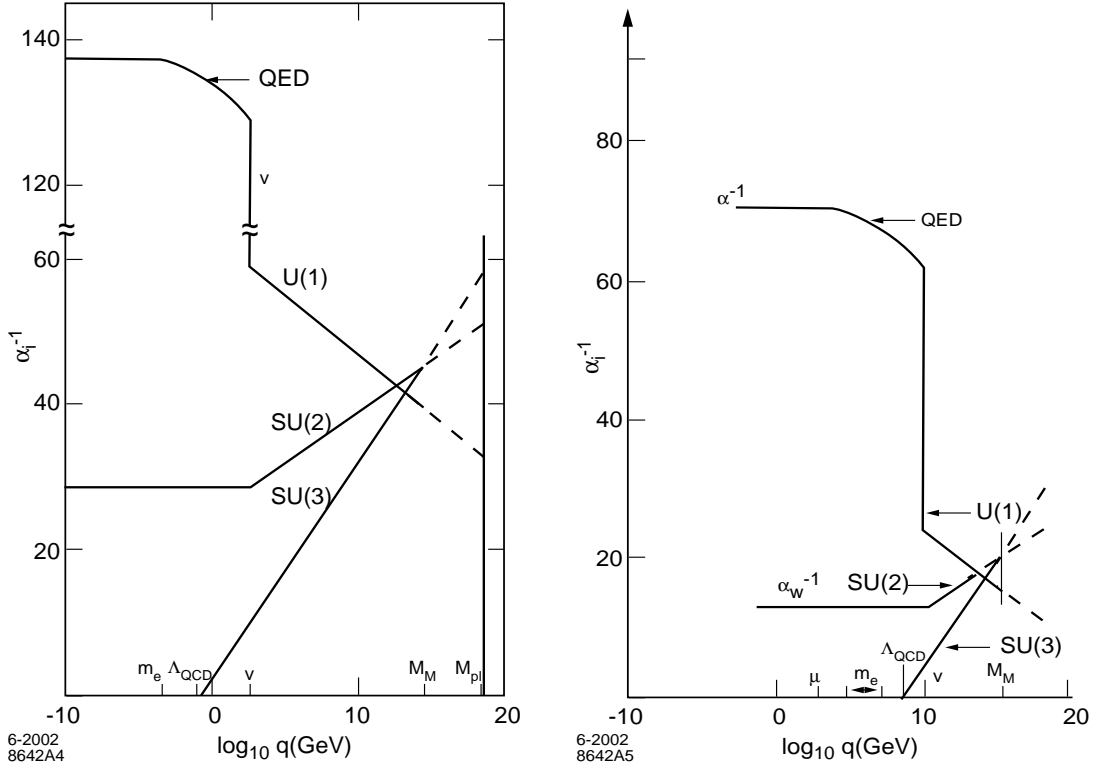

\centerline{
\includegraphics*[width=2.75in,height=4in]{8642A04}
\hspace*{3mm}
\includegraphics*[width=2.75in,height=4in]{8642A05}}
\caption[*]{\baselineskip=14pt Running of the coupling constants
for (a) our universe, and for (b) a universe with horizon size of
10$^{-2}$ cm. \label{fig:3}}
\end{figure}

\begin{figure}[htb]
\centering
\includegraphics{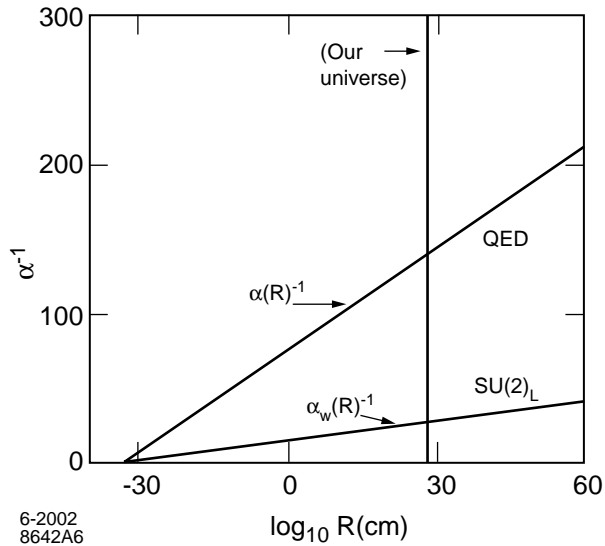}
\caption[*]{\baselineskip=14pt Dependence of $\alpha$ and
$\alpha_{\rm weak}$ (evaluated in the infrared) versus $\log_{10}
R$. \label{fig:4}}
\end{figure}

It is important that this behavior holds for $\alpha_s$ evaluated
at the GUT scale $M$. As long as coupling constant unification
makes any sense at all, we may infer that the electroweak and
electromagnetic couplings must also possess the same behavior.
Since
\begin{equation}
\alpha^{-1}_s(M^2, R^2) = b_s\ \ell n\,
\frac{M^2}{\Lambda^2_{QCD}} \approx \alpha^{-1}_1(M^2, R^2)\simeq
\alpha^{-1}_2(M^2, R^2) \label{eq:abc}
\end{equation}
it follows that the electroweak couplings at the weak scale are
\begin{equation}
\alpha^{-1}_i(v^2,R^2) = \alpha^{-1}_i(M^2,R^2) + b_i\ \ell n\,
\frac{M^2}{v^2} = b_s\ \ell n\, \frac{M^2}{\Lambda^2_{QCD}} + b_i\
\ell n\, \frac{M^2}{v^2}\ . \label{eq:def}
\end{equation}
But because $M/v$ and $M/\Lambda_{QCD}$ scale as powers of
$M_{\\pl}R$, it follows that
\begin{equation}
\alpha^{-1}_i(v^2,R^2) = ({\rm const}) \, \ell n\, M_{pl}R \qquad
\qquad i = 1,2 \label{eq:ghi}
\end{equation}
and
\begin{equation}
\frac{\alpha_1(v^2,R^2)}{\alpha_2(v^2,R^2)} \equiv \tan^2\theta_W
= {\rm const} \ . \label{jkl}
\end{equation}
Therefore the weak mixing angle $\theta_W$ is to good
approximation independent of $R$, as is the ratio
$\alpha/\alpha_W\cong \alpha(R^2)/\alpha_2(v^2,R^2)$. The usual
diagram of gauge coupling running and unification is shifted in
scale as $R$ is varied. Because the weak mixing angle does not
depend upon $R$, the entire figure becomes self-similar
(Fig.~\ref{fig:3}). To leading order, 1/$\alpha$ and
1/$\alpha_{\rm weak}$ depend linearly on $\ell n\, R$ and vanish
at the Planck/GUT radius (Fig.~\ref{fig:4}).

Finally, we shall again invoke Occam and assume that the large
Higgs Yukawa couplings $\lambda$ and $h^2_{\rm top}$ are no
exception to the rule, and that they also obey the same rule,
namely that the inverse couplings vary linearly with $\ell n\, R$
and vanish for $R$ at the Planck/GUT scale. The usual
renormalization group equations connecting the Higgs self-coupling
$\lambda$ to the top-quark Higgs coupling $h_{\rm top}$ (with
important QCD corrections) remain unchanged. The new equations are
again
\begin{equation}
R^2 \, \frac{\partial}{\partial R^2}\,
\left(\frac{1}{h^2_t}\right) = \mathrm{constant} \qquad R^2\,
\frac{\partial}{\partial R^2}\, \left(\frac{1}{\lambda}\right) =
\mathrm{constant}\label{eq:m}
\end{equation}
and may be used to determine how $\lambda$ and $h_{\rm top}$,
evaluated at either the GUT scale or the infrared scale, vary as
$R$ is varied.

We now return to consideration of other standard-model parameters
with dimension of mass, starting with the masses of top quark,
Higgs bosons, and electroweak gauge bosons, all of which have a
mass formula of the form
\begin{equation}
m = gv \label{eq:n}
\end{equation}
where $g$ stands for a generic dimensionless coupling constant.
This implies a renor\-mali\-zation-group equation of the form
\begin{equation}
\frac{R}{m}\, \frac{\partial m}{\partial R} =
\frac{R}{v}\, \frac{\partial v}{\partial R} +
\frac{R}{g}\, \frac{\partial g}{\partial R}=
p_v + \mathrm{(constant)} \cdot g^2
\ \label{eq:o}
\end{equation}
which clearly possesses an order $g^2$ ``radiative correction" to
the leading behavior. What is clearly happening is that the ratio
$m/v$ is stable, and does not run as a power of $R$ but only as a
power of $g$, {\em i.e} of $\ell n \, R$.

In the case of these particles the corrections are not very
important, because their masses are so close to the value of the
electroweak vev $v$. A more dramatic example is given by the electron
mass, which in a sense lies at the opposite extreme. We do not
know whether to regard the ratio $m_e/v$ as a function of
dimensionless coupling constants, {\em i.e.} dependent only on
$\ell n\, R$, or as a ratio of fundamental scales, {\em i.e.}
dependent on a power of $R$. In the former case we have
\begin{equation}
\frac{v}{m_e} \sim (\ell n\, MR)^n \label{eq:p}
\end{equation}
where $M$ is the GUT/Planck mass scale.  In the latter case the
flow is
\begin{equation}
\frac{v}{m_e} \sim (MR)^{p_e} \qquad p_e \approx 0.1 \ .
\label{eq:q}
\end{equation}
For numerical estimation of the former case we shall choose $n =
4$ for the electron. Our motivation is simply to assign one power
of some $g^2$, of typical order of magnitude, {\em i.e.} $(\ell n\,
MR)^{-1}$, per mass hierarchy level. Thus for bottom, strange,
down, electron, we take $n = 1,2,3,4$ respectively.

Returning to the question of $R$-dependence of electron mass, we
can evaluate each case. The results are shown in Fig.~\ref{fig:5},
and we may regard the shaded region as a one-parameter region of
uncertainty. We also plot in Fig.~\ref{fig:6} the $R$-dependence
of the ratio of electron mass to proton mass, which is a crucial
parameter for chemistry and condensed matter physics.

\begin{figure}[htbp]
\centering
\includegraphics{8642A07}
\caption[*]{\baselineskip=14pt Dependence of the electron mass,
scaled to the electroweak vev $v$, versus $\log_{10} R$. The
shaded region is a region of uncertainty. \label{fig:5}}
\bigskip
\centering
\includegraphics{8642A08}
\caption[*]{\baselineskip=14pt Dependence of $(m_e/m_p)$ on
$\log_{10} R$. \label{fig:6}}
\end{figure}

The $R$-dependence of other small masses present in the standard
model should be similarly regarded, especially the up and down
quark masses which drive chiral symmetry breaking of the strong
interactions and are responsible for the pion mass. For the strong
interactions there is to good approximation only the scale
$\Lambda_{QCD}$, which by itself determines all masses other than
that of the pions. In particular the proton mass is proportional
to $\Lambda_{QCD}$, as well as the masses of all mesons and
baryons other than the pions  (and kaons). But the scale set by
the pion mass, whose square varies linearly with the light quark
masses and with $\Lambda_{QCD}$, does matter. It is the $R$
dependence of the ratio of pion to proton mass which will be the
crucial parameter for nuclear physics. Its square is plotted in
Fig.~\ref{fig:7}. In Fig.~\ref{fig:8}, we also plot $(m_K/m_\pi)^2$,
which is proportional to the
ratio of strange-quark to down-quark mass.

\begin{figure}[htbp]
\centering
\includegraphics{8642A09}
\caption[*]{\baselineskip=14pt Dependence of $(m_\pi/m_p)^2$ on
$\log_{10} R$. \label{fig:7}}
\bigskip
\centering
\includegraphics{8642A10}
\caption[*]{\baselineskip=14pt Dependence of $(m_K/m_\pi)^2$ on
$\log_{10} R$ \label{fig:8}}
\end{figure}

The remaining parameters of the standard model are $\theta$, the
$CP$-violating parameter within QCD, and the CKM mixing
parameters, which are closely related to the small quark and
lepton masses. In addition there are neutrino masses and mixings.
The underlying physics still awaits better understanding, and we
have little to add here. These parameters do not appear to be of
great importance for what follows in the remainder of this note.

\section{Properties of Matter in the Ensemble of
Universes}\label{sec:3}

One of our main goals is to investigate the properties of
universes assumed to be almost the same as ours, {\em i.e.} with
radii $R$ almost the same as ours, and with cosmological initial
conditions similar to ours. However we shall also consider more
extreme cases, namely universes with radii within 30 orders of
magnitude of our own. One reason for considering such large
bandwidth is that there will be some properties of these universes
which are robust, and do not vary all that much over all those
powers of 10. For example the weak and electromagnetic
fine-structure constants are in this context robust, having values
in this range of $R$ which are within a factor two of what we
observe. On the other hand it is well known, especially amongst
the ``anthropic" community, that other properties of our universe
are very finely tuned and will only exist over a quite small
bandwidth. We shall pay special attention to such ``anthropic"
constraints, as discussed for example in the book by Barrow and
Tipler \cite{ref:d}, and will be interested in the bandwidth in
$R$ for which they are satisfied.

We shall begin by considering how the properties of elementary
particles, of nuclear matter, and of ordinary matter vary with
$R$. We then investigate how the structure of astrophysical
objects of interest, such as planets, stars, {\em etc.} vary as
$R$ is varied.

\subsection{Elementary Particle Properties}
\label{ssec:a}

Even for the smallest universe that we shall consider, with radius
100 microns, there is good separation (a factor 10--20) between
the electroweak scale and the strong interaction scale. Heavy
quarks, electroweak gauge bosons, Higgs particles, etc. are still
unstable and will not grossly influence the phenomenology of
ordinary matter. A marginal case is that of the strange quark. An
estimate of its effect is given by the ratio of kaon to pion mass
exhibited in Fig.~\ref{fig:8}. We see that even for the extreme
cases the ratios always stay comfortably above unity, suggesting
that we do not err badly in neglecting strange-quark contributions
to ordinary matter. The strange hadron masses appear to stay high
enough to allow semileptonic weak decays at the very least to
proceed.

\subsection{Nuclear and Atomic Matter}
\label{ssec:b}

Crucial to the properties of nuclear and atomic matter are the
values of the fine-structure constant (here constrained to a
reasonable range of values), the ratio of electron to proton mass,
the ratio of pion to proton mass, and the neutron-proton mass
difference. As long as the electron-proton mass ratio stays small,
atomic physics and chemistry will remain recognizable. We see from
Fig.~\ref{fig:6} that this is in fact the case. Likewise, in
Fig.~\ref{fig:7} we see that the pion mass stays well below the
proton mass over all the range to be considered.

We conclude that over the 60 orders of magnitude we shall
consider, it would appear that chemistry and condensed matter
physics will be at least existent and reasonably recognizable. As
for nuclear matter, it should exist in recognizable form as the
pure chiral limit is approached, because the long range force due
to pion exchange is not crucial.  It might bind the nucleons a
little more (or less), but probably not enough to change the phase
structure. However as the pion mass increases, there is more
potential for trouble. According to Fig.~\ref{fig:7}, this appears
to occur only in the largest or smallest universes that we shall
consider.  All this will be discussed in more detail in Section
\ref{ssec:d}.

\begin{figure}[htb]
\centering
\includegraphics{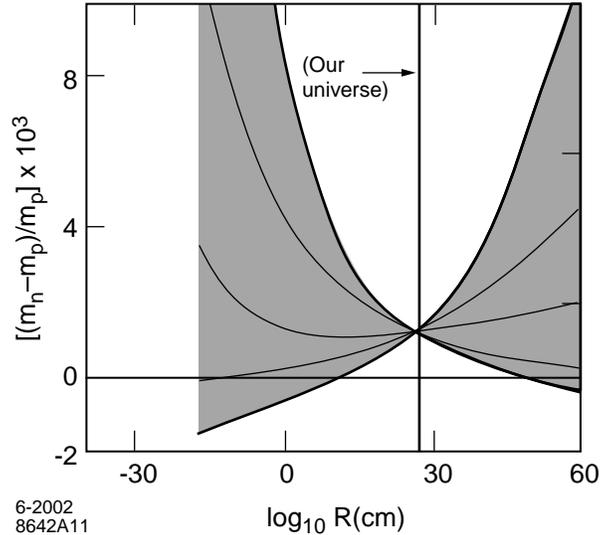}
\caption[*]{\baselineskip=14pt Dependence of the neutron-proton
mass difference $\Delta m$, scaled to the proton mass, on
$\log_{10} R$. \label{fig:9}}
\end{figure}

On the other hand, as the pion mass decreases, the neutron-proton
mass difference varies in a nontrivial way. It is composed of two
pieces. The dominant one is due to the mass difference of the up
and down (current-) quarks, and the other is electromagnetic \cite{ref:l}.
Schematically we may write
\begin{eqnarray}
\Delta m = (m_n-m_p) &=& a_1\, (m_u-m_d) + b(\alpha\, m_p)
\nonumber
\\[1ex]
&=& a_2 \, \left(\frac{m_u-m_d}{m_u+m_d}\right)\,
\frac{m^2_\pi}{m_p} + b(\alpha\, m_p) \nonumber \\[1ex]
&=& a_3\, \frac{m^2_\pi}{m_p} + b(\alpha\, m_p) \\[1ex]
&\cong& 0.1\ m_p \left[\left(\frac{m_\pi}{m_p}\right)^2-
\alpha\right] \nonumber \label{eq:r}
\end{eqnarray}
where we assume that $a_i$, $b$, and the ratio of the difference
of up and down quark masses to their sum are to good approximation
scale-independent. In the last line we have used the accepted
values of the two contributions \cite{ref:l} in approximate form
to provide a useful mnemonic. Note that the electromagnetic and
quark contributions to $\Delta m$ are of opposite sign. As the
chiral limit is approached, the neutron becomes stable and the
proton unstable. The latter case is clearly a serious matter for
atomic physics and chemistry, which might even cease to exist.
However, from Fig.~\ref{fig:9} we see that the only cases where
this becomes a problem are for universes whose radii are ten to
fifteen orders of magnitude larger or smaller than the radius of
our universe. The cosmologies for those cases will evidently be
nontrivially different from our own, and we will briefly return to
this issue later.

\subsection{Stable Cosmological Objects}\label{ssec:c}

Other than black holes, all the large stable cosmological objects
exist as a consequence of the Pauli principle. Fermion degeneracy
pressure in one form or another provides the repulsion that
prevents such objects to gravitationally collapse. This mechanism
is so robust that we can expect it to operate over the whole 60
orders of magnitude of radii $R$ which we consider. Three obvious
classes to consider are planets, white dwarfs, and neutron stars.
In these three cases the degeneracy pressure is provided by
nonrelativistic electrons, relativistic electrons, and neutrons
respectively. We begin by briefly reviewing these cases.

The density of a planet-like object is fixed by the interatomic
force, and the spacing of atomic nuclei is of order $(\alpha
m_e)^{-1}$. This gives for the baryon number of a planet of radius
$r$ the value
\begin{equation}
B \cong A(\alpha\, m_e r)^3 \label{eq:s}
\end{equation}
where $A$ is the mean atomic number of the nucleus. (If heavy
elements are not produced in the universe of interest, then we
take $A = 1$, and limit our attention to Jupiter-like planets).
The chemical binding energy per nucleus is of order the Rydberg,
and from this we can determine the total chemical energy and
equate it with the gravitational energy in order to determine the
characteristic size $r$ of the planet:
\begin{equation}
U_{\mathrm{chem}} \sim (\alpha^2m_e)\, \frac{B}{A} \sim
\frac{B^2}{r}\, \left(\frac{m_p}{M_{pl}}\right)^2 \sim
U_{\mathrm{grav}} \ . \label{eq:t}
\end{equation}
Upon eliminating $r$, this leads to
\begin{equation}
B \sim \frac{\alpha^{3/2}}{A^2}\ \left(\frac{M_{pl}}{m_p}\right)^3
\end{equation}
In a similar way, we may consider white dwarfs, where relativistic
electron degeneracy pressure balances the gravitational energy. In
that case the baryon number is given by
\begin{equation}
B \sim r^3\, p^3_F    \label{eq:u} \label{eq:v}
\end{equation}
where $p_F$ is the Fermi momentum of the electron plasma.
The energy-balance equation is
\begin{equation}
U_{\mathrm{degen}} \sim p_F\, B \sim \frac{B^2}{r}\,
\left(\frac{m_p}{M_{pl}}\right)^2 \sim U_{\mathrm{grav}}
\label{eq:w}
\end{equation}
which simplifies to
\begin{equation}
B \sim \left(\frac{M_{pl}}{m_p}\right)^ 3 \ . \label{eq:x}
\end{equation}
Finally, we may consider the case of the neutron star. It is
similar to the white dwarf case. One simply replaces the electron Fermi
momentum $p_F$ with $\Lambda_{QCD}$ which characterizes the neutron
Fermi momentum, and arrives at the same result.

\begin{figure}[htb]
\centering
\includegraphics{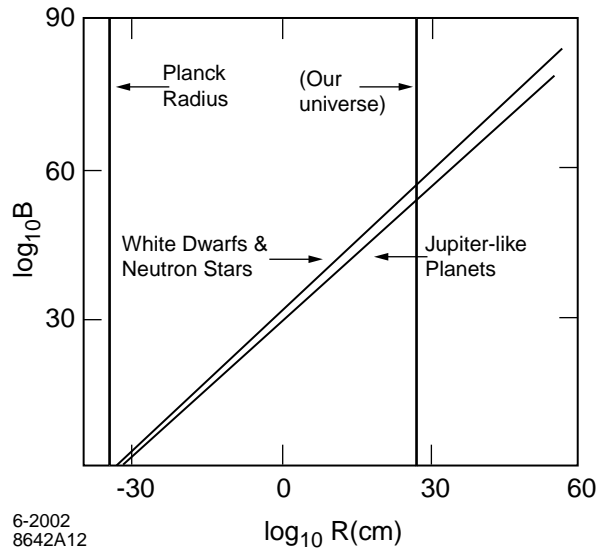}
\caption[*]{\baselineskip=14pt Dependence of baryon number $B$ of
astrophysical objects upon $\log_{10} R$. \label{fig:10}}
\end{figure}

We see that in all three cases the baryon number, hence the mass,
of the object scales as the inverse third power of the proton
mass, and therefore scales as the appropriate power of $R$. The
result is shown in Fig.~\ref{fig:10}. We therefore can anticipate
the existence and can understand the properties of these massive
objects, throughout the 60 orders of magnitude of $R$ we consider.
Stars, however, are another matter. The question of whether these
large objects ignite and burn, and for how long, depends on
details. Before addressing stellar structure we consider some of
the finer points having to do with the nuclear force.

\subsection{The Nuclear Force}\label{ssec:d}

The simplest system in nuclear physics is the dinucleon. It is a
delicate case, especially in the context of astrophysics, because
the nonexistence of bound diprotons and/or dineutrons is needed to
keep stars burning \cite{ref:d,ref:o}, and the existence of a
deuterium bound state is an essential ingredient for fusion
reactions in stars as well as in  big-bang nucleosynthesis. What
is needed is the dependence of the binding energies upon
$(m_\pi/m_p)$. There is no consensus on what the answer is.
Depending upon the method, different magnitudes and even signs are
obtained \cite{ref:p,ref:q,ref:r}.   What is important in our
application is the value of $(m_\pi/m_p)$ for which the deuteron
becomes unbound, as well as the value of $(m_\pi/m_p)$ for which
the diproton and/or dineutron might become bound.  We choose here
an estimate which lies in the midrange of what is generally
considered \cite{ref:r,ref:s}, and has the sign dictated by naive
intuition; as one approaches the chiral limit, the binding
energies increase. Our choices are as follows:
\begin{eqnarray}
\mathrm{Deuteron\ bound\ if\ } m_\pi/m_p &\leq& 0.16 \nonumber \\
\mathrm{Diproton\ bound\ if\ } m_\pi/m_p &\leq& 0.08 \ .
\label{eq:y}
\end{eqnarray}
We emphasize that these choices are uncertain, but probably by not
more than a factor 3. However, it is arguable \cite{ref:r,ref:s}
that the dinucleon remains unbound even in the chiral limit, in
contradiction to the choice made above. But, it will turn out that
in what follows we will not consider any region of parameter space
where, given the parameters we have chosen, the diproton is bound,
so that for us the issue is moot.

Finally, we may consider the mechanism for producing carbon in
stars. This depends upon the existence of the anthropically famous
triple-$\alpha$ reaction \cite{ref:xxx}
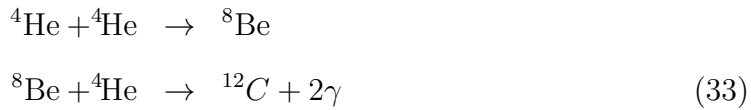
\begin{eqnarray}
 ^4\!\,\mathrm{He} + ^4\!\!\mathrm{He}&\rightarrow&
 ^8\!\,\mathrm{Be}\nonumber \\
 ^8\!\,\mathrm{Be} + ^4\!\!\mathrm{He} &\rightarrow& ^{12}C +
 2\gamma
 \label{eq:z}
\end{eqnarray}
with the resonance in $^{12}C$ predicted by Hoyle \cite{ref:u},
together with the absence of a crucial level in $^{16}O$. The
parameter sensitivity of this process, which is of order $\delta E
\sim$ 100 keV in a system with binding energy scales in the 10 MeV
range, is discussed by Oberhummer {\em et al.} \cite{ref:v}, among
others \cite{ref:w}. The result is that an 0.3 percent variation
in the overall strength of the nuclear force is enough to strongly
modify  this delicately balanced mechanism. If such a perturbation
were applied to the deuteron, it would change its binding energy
by about five percent. We conclude that at most the
triple-$\alpha$ process represents a parameter sensitivity a
factor 20 greater than what one obtains from considering dinucleon
binding. However, the actual sensitivity of the triple alpha
mechanism may be considerably less, because the dinucleon binding
could be more sensitive to the long-range pionic tail of the force
than the interactions between compact, closed-shell
alpha-particles.  It would be helpful to have a good description
of the dependence of the intermediate-range, isosinglet,
spin-independent attractive force upon pion mass.  But at present
this seems not to exist.

\section{Cosmology}\label{sec:4}

In this section we explore how big bang cosmological evolution
depends on the ultimate ``size" $R$ of the particular universe
which is created. There are a variety of epochs in the history of
a universe which are especially sensitive to parameter variations.
Before going into more details we briefly sketch them here to set
the stage:

1. We take as initial condition of the universe its state just
after inflationary reheating (the assumption that inflation indeed
occurs will not be too important), with the initial temperature
taken to be of the Planck/GUT scale. The universe is always taken
to be spatially flat. The magnitude of the primordial density
fluctuations, which eventually account for the observed
fluctuation spectrum in the 3 degree microwave background, is in
principle a parameter to be specified. In practice we shall choose
it to be equal to what it is in our universe, $\delta_H =
\delta\rho/\rho \sim 2 \times 10^{-5}$, independent of $R$.

\begin{figure}[htb]
\centering
\includegraphics{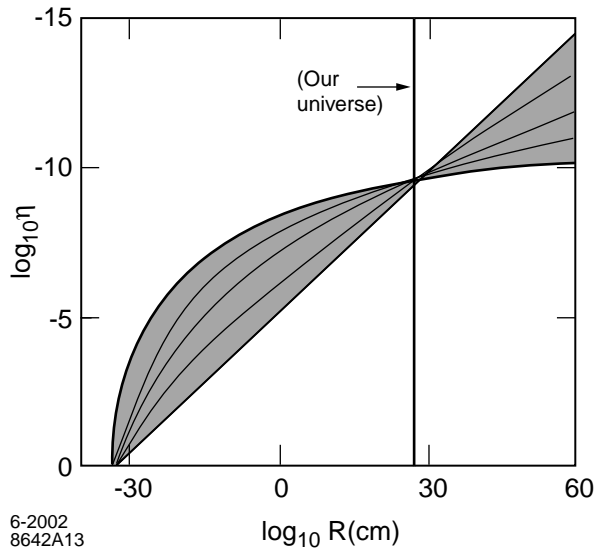}
\caption[*]{\baselineskip=14pt Conjectured dependence of the
baryon-to-photon entropy ratio $\eta$ upon $\log_{10} R$.
\label{fig:11}}
\end{figure}

2.  The baryon asymmetry of the universe is assumed to be
generated in some intrinsic way from unknown,
extended-standard-model mechanisms at a very high temperature
scale. The details of this mechanism are at present very
uncertain. Therefore the $R$ dependence of this asymmetry will be
treated in a way similar to how the electron mass was treated. We
assume that for universes with a size of order the Planck radius,
the baryon asymmetry is large, of order unity. We assume that the
interpolation from the Planck size to large universes like our own
may behave as a power of $R$, or as a power of $\ell n\, R$, each
option taken to be an extreme case. The result is shown in
Fig.~\ref{fig:11}. While the uncertainties become large for
universes very different in size from our own, at least the
dependence upon $R$ is monotonic.

3.  As the universe cools, the phase transitions at electroweak
and QCD scales proceed in a way similar to our universe. (The
baryon asymmetry may be modified at the electroweak scale via
``sphaleron" effects \cite{ref:x}, and if so it is the modified
asymmetry which is shown in Fig.~\ref{fig:11}.) Differences appear
at the epoch of nucleosynthesis, at a temperature of order
$10^{-3}m_p$. The mechanisms are sensitive to the time at which
neutrinos decouple from the plasma and baryonic chemical
equilibrium is lost. The abundances of $^4$He, deuterium, and
hydrogen (and even $^2$He) become sensitive to the parameters and
require a detailed discussion.

4.  At some very uncertain temperature scale, (cold) dark matter
decouples from the plasma and evolves, eventually becoming a major
component of the matter density. The physics of this is obscure.
We shall assume that the cold dark matter is composed of WIMPs, by
which we mean that their interactions with each other and with
ordinary matter are characterized by a scale somewhere around the
electroweak scale.

5.  The epochs of matter-radiation equality and of decoupling of
radiation from matter are also parameter sensitive. What happens
during these periods provide initial conditions for the subsequent
evolution of large-scale structure formation. All of this will
require a detailed discussion.

\subsection{Nucleosynthesis}\label{ssec:aa}

\begin{figure}[htb]
\centering
\includegraphics{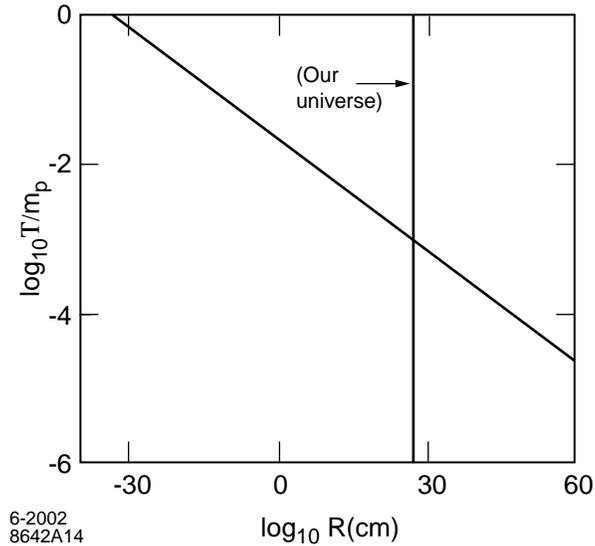}
\caption[*]{\baselineskip=14pt Dependence of the temperature at
which neutrinos decouple from matter, scaled to the proton mass,
upon $\log_{10} R.$ \label{fig:12}}
\end{figure}

As the universe cools below the QCD phase transition, quarks and
antiquarks bind into mesons and baryons, and the mesons soon
disappear. Neutrons and protons are kept in chemical equilibrium
by electroweak scattering processes induced by neutrinos.
Eventually the neutrinos decouple, the criterion for decoupling
being that the expansion rate of the universe exceed the collision
rate. The expansion rate for a radiation dominated universe is
\begin{equation}
H^2 \sim \frac{T^4}{M^2_{pl}} \ . \label{eq:aa}
\end{equation}
Equating $H$ to the collision rate gives
\begin{equation}
H \sim \frac{T^5}{v^4} \ . \label{eq:bbb}
\end{equation}
This leads to the criterion
\begin{equation}
\left(\frac{T}{m_p}\right) \sim \left(\frac{v}{m_p}\right)\
\left(\frac{v}{M_{pl}}\right)^{1/3} \ . \label{eq:cc}
\end{equation}
This result is plotted in Fig.~\ref{fig:12}.

In our universe, neutrino decoupling occurs at a temperature of
about 0.7 MeV. At that temperature the ratio of neutrons to
protons has been depleted by about a factor seven, due to the
Boltzmann factor containing the neutron-proton mass difference.
The remaining neutrons capture into deuterium, which is then
converted quickly to $^4$He by fusion reactions. The net result is
a primordial helium abundance of 22 percent or so.

Had neutrino decoupling occurred much earlier, the neutron-proton
ratio would have been unity. All the baryons would end up as
deuterium, which would then convert via fusion to helium.
Conversely, if decoupling were to occur much later, then the
neutrons would be removed by the neutrino reactions, and there
would be nothing left at low temperatures but hydrogen. We
therefore expect the dependence of the abundance of primordial
helium to change from very high for $R$ smaller than the radius of
our universe to very low for $R$ greater.

\begin{figure}[htb]
\centering
\includegraphics{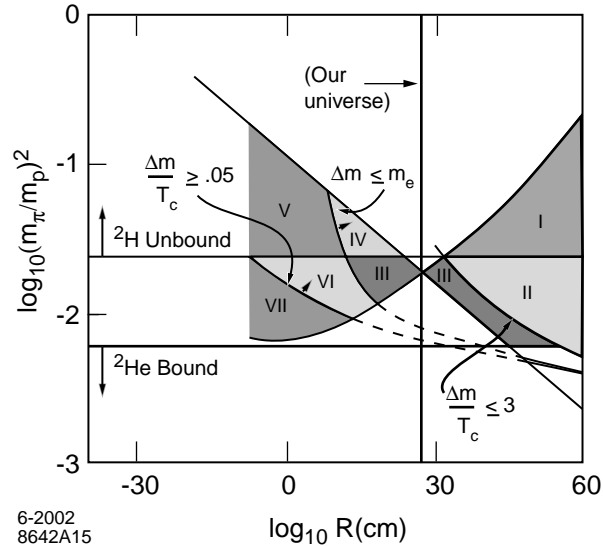}
\caption[*]{\baselineskip=14pt Regions of parameter space for
which cosmological evolution is qualitatively different. See the
text for the details.  \label{fig:13}}
\end{figure}

However, for very large or small $R$ the situation is more
complicated and in fact uncertain. As $R$ varies the ratio of pion
to proton mass varies, and with it the binding energy of the
dibaryons. For large values of $(m_\pi/m_p)$, the deuteron
probably does not exist, and the fusion reactions are blocked. If
the pion mass is very small, then the diproton may be bound, and
instead of hydrogen in the final state of proton universes, there
would initially be $^2$He. Further complicating the situation is
the $R$ dependence of the ratio of neutron-proton mass difference
to electron mass, which for small $R$ can fall below unity,
leading to a stable neutron. In addition, if $(m_\pi/m_p)$ is
sufficiently small, the neutron-proton mass difference changes
sign. The situation is sketched out in Fig.~\ref{fig:13}, where we
identify various regions in the two-dimensional parameter space of
$R$ and $(m_\pi/m_p)$ for which the baryogenesis scenarios
qualitatively change. There are seven distinct regions of the
parameter space we consider. Regions I, IV, and V are
characterized by an unstable deuteron. Regions V, VI, and VII are
characterized by a stable neutron. In most of region VII the
proton is unstable. In regions I and II neutrinos decouple at such
a low temperature that the fraction of neutrons in the mix is less
than 5 percent. Consequently those universes evolves into
predominantly hydrogen. In region VII the opposite occurs, and the
$^4$He fraction exceeds 90 percent. Only in region III is the
situation qualitatively the same as for our universe. If $R$ lies
between $10^{24}$ cm  and $10^{33}$ cm, this is assured to be the
case, although this conclusion rests heavily upon the assumption
made in Eq. (\ref{eq:y}).

We now briefly describe the individual baryogenesis histories for
the seven regions we have identified:

Region I: In this region the deuteron is unbound, and the
decoupling of neutrinos occurs so late that the $n/p$ ratio is
less than 5 percent. The result is a nearly pure hydrogen
universe. However fusion reactions within stars will not proceed
because of the absence of deuterium.

Region II: Again the $n/p$ ratio is less than 5 percent, and a
predominantly hydrogen universe is formed. But now the deuteron
exists, so that in principle stars can burn hydrogen into helium.

Region III: As noted above, this region resembles---and
includes---our own universe.

Region IV:  In this region the deuteron is unbound. Although
neutrinos decouple relatively early, when the $n/p$ ratio is not
too small, the fusion reactions are blocked. The extra neutrons
decay, and we are again left with a hydrogen-dominated universe.
But as in Region I, fusion reactions within stars are blocked.

Region V:  This region differs from Region IV, because here the
neutron is stable. The universe will be mixed hydrogenic and
neutron, with fusion reactions again blocked because of the
absence of deuterium.

Region VI:  In this region, deuterium exists. The $n/p$ ratio at
neutrino decoupling is not small, so that nucleosynthesis of
$^4$He should proceed. The neutron is stable, but primordial
neutrons are presumably found in the helium. Fusion reactions in
stars should be able to proceed.

Region VII: In most of this region the proton is unstable and
decays to the neutron with positron emission. The $p/n$ ratio is
large enough (but less than unity!) so that primordial $^4$He will
be produced. Because the deuteron is stable, fusion reactions in
stars may proceed. Hydrogen-based chemistry will not exist,
although perhaps some deuterium-based chemistry might survive.

It is noteworthy that in all seven regions the electron chemical
potential does not vanish.  An electron plasma will persist until
decoupling occurs at a much lower temperature scale.

\subsection{Stars}\label{ssec:bb}

The properties of these regions are perhaps well enough defined
that one could go further and map out the subsequent cosmological
history in a little more detail. We shall not try to do so here.
But before going on to more general cosmological questions, we
will consider (within our Region III) the additional oft-cited
constraints on the existence of long-lived stars and of the
conditions appropriate to the production of carbon and other
heavier elements \cite{ref:o,ref:d,ref:e}. An immediate reason for
doing so is anthropic; we would like to know the bandwidth in $R$
within which the changes in standard model parameters are small
enough to preserve the conditions in our universe which are
conducive to life as we know it.

Quite a long list of ``anthropic" constraints exist. Upon
examination of the items on that list, it should come as no
surprise that the most restrictive by far is the existence of the
triple-$\alpha$ fusion-reaction chain which allows the production
of carbon and thereby the existence of heavier elements. We
already mentioned in Section \ref{sec:3} that this constraint
could be 20 times more sensitive than the constraints used above
regarding the existence of bound deuterium. Examination of
Fig.~\ref{fig:13} shows that this enhanced sensitivity roughly
translates into
\begin{equation}
\left|\log_{10}\frac{R}{R_0}\right| < 0.2 \label{eq:dd}
\end{equation}
or that
\begin{equation}
0.7 < \frac{R}{R_0} < 1.5 \label{eq:ee}
\end{equation}
where $R_0$ is the radius of our universe. In other words, if the
radius of a universe in our ensemble of universes is within
roughly a factor $\sqrt2$ of ours, the standard model parameters
are close enough to our own not to upset the conditions necessary
for existence of life in that universe.

\subsection{Large Scale Structure}\label{ssec:cc}

As the universe continues to cool and expand, the era of matter
dominance emerges. In the scenario we consider ($\Lambda CDM$), it
is the cold dark matter that is essential in initiating the growth
of density fluctuations. The baryons carry less of the energy
density, and they stay coupled to the photons for much longer,
thereby being unable to fully participate in the growth of
inhomogeneities until decoupling is reached.

We review briefly the standard calculations in order to see the
parameter dependencies \cite{ref:x}. The abundance of cold dark
matter WIMP particles $X$ is estimated by equating their rate of
production and/or annihilation to the Hubble expansion rate at the
time/temperature of WIMP decoupling.
\begin{equation}
n_X \VEV{\sigma v} \sim H \sim \frac{T^2}{M_{pl}} \ .
\label{eq:ff}
\end{equation}
Here $n_X$ is the number density and $T$ the temperature at
decoupling. Normalizing the abundance to the abundance of photons,
proportional to the cube of the temperature, gives
\begin{equation}
\left(\frac{n_X}{n_\gamma}\right) \sim \frac{1}{M_{pl}T\VEV{\sigma
v}} \sim \frac{20}{m_X M_{pl}\VEV{\sigma v}} \label{eq:gg}
\end{equation}
where we use the fact that within a factor two the WIMP decoupling
temperature is twenty times lower than the rest mass of the $X$
particles over a very wide range of parameters \cite{ref:x}.

We may further relate this to the abundance of baryons, by
introducing the baryon-to-photon entropy ratio $\eta$, already
discussed above and depicted in Fig.~\ref{fig:11}:
\begin{equation}
\frac{\Omega_X}{\Omega_B} \sim \left(\frac{m_X}{m_p}\right)\
\left(\frac{n_X}{n_\gamma}\right) \
\left(\frac{n_\gamma}{n_p}\right) \sim \frac{20}{m_p
M_{pl}\VEV{\sigma v}\eta } \ . \label{eq:hh}
\end{equation}
When this is evaluated for our universe, the cross section
estimate is
\begin{equation}
\VEV{\sigma v} \sim \VEV\sigma \sim 10^{-37} \mathrm{cm}^2 \ .
\label{eq:ii}
\end{equation}
This cross section is close to the electroweak scale. Therefore we
assume, as do many others \cite{ref:cc}, that it scales with the
inverse square of the electroweak vev $v$ and perhaps some power
of a coupling constant,
\begin{equation}
\VEV{\sigma v} \sim \frac{\alpha^n}{v^2} \label{eq:jj}
\end{equation}
with $n$ taken to be two or three.

\begin{figure}[htb]
\centering
\includegraphics{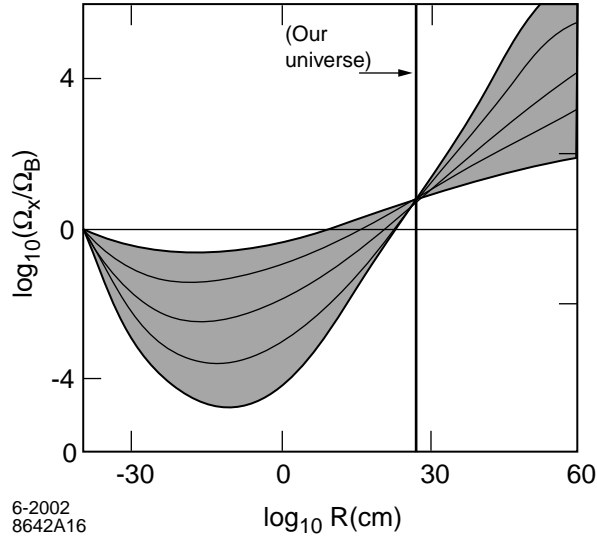}
\caption[*]{\baselineskip=14pt Estimated ratio of dark matter to
baryonic matter $\Omega_X/\Omega_B$ as a function of $\log_{10}
R$. \label{fig:14}}
\end{figure}

We may now look at the $R$-dependence of the ratio of dark matter
to baryonic matter. It is plotted in Fig.~\ref{fig:14}. We see
that dark matter will dominate over baryonic matter provided the
radius of the universe is greater than $10^{-4}$ of our own.

\begin{figure}[htb]
\centering
\includegraphics{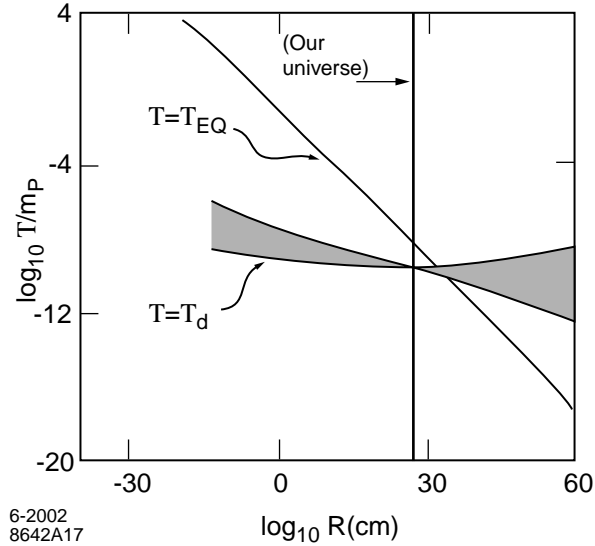}
\caption[*]{\baselineskip=14pt Temperature $T_{EQ}$ for which
nonrelativistic matter and radiation are equal as a function of
$\log_{10} R$. Also shown is the temperature $T_d$ at which matter
and radiation decouple, versus $\log_{10} R$.  \label{fig:15}}
\end{figure}

We may also determine the temperature $T_{EQ}$ when the
contributions of matter and radiation are equal, and which signals
the onset of matter-dominated expansion of the universe. From Eq.
(\ref{eq:gg}) we find
\begin{equation}
\frac{\rho_X}{\rho_\gamma} \sim \frac{m_X}{T}\
\left(\frac{n_X}{n_\gamma}\right) \sim \frac{20}{T\, M_{pl}
\VEV{\sigma v}} \sim \frac{20\, v^2}{T\, M_{pl}\,\alpha^n}
\label{eq:kk}
\end{equation}
from which it follows
\begin{equation}
T_{EQ} \sim 20\ \frac{v^2}{M_{pl}\,\alpha^n} \ . \label{eq:ll}
\end{equation}
This is plotted in Fig.~\ref{fig:15}. Also shown there is the
temperature at which radiation decouples from ordinary matter and
the universe becomes transparent. The formula which controls this
is \cite{ref:x}
\begin{equation}
T_d \sim \frac{\alpha^2m_e}{40} \ . \label{eq:mm}
\end{equation}
We see that if $R$ is less than $10^5$ the size of our universe
\begin{equation}
T_d < T_{EQ} \  \label{eq:nn}
\end{equation}
while the opposite is true for larger universes.  Therefore for
small universes the WIMP degrees of freedom will first feel the
Jeans instability, with the growth of fluctuations in baryonic
matter occurring later.  For the large universes the baryonic
fluctuations grow together with the WIMP fluctuations.  It is not
immediately clear how much of a difference this might make in the
creation of large-scale structure.

The evolution of the large scale structure is in general a complex
topic. The most straightforward part of the subject consists in
the growth of small density perturbations in the linear regime. As
mentioned above, we assume that the typical scale of primordial
perturbations, present at the earliest epoch we consider when
temperatures were at the Planck/GUT scale, are of order
\begin{equation}
\left(\frac{\delta \rho}{\rho}\right)_0 \sim 2 \times 10^{-5}
\label{eq:oo}
\end{equation}
as measured by the cosmic microwave background temperature
fluctuations. These density perturbations remain frozen at more or
less this value during the radiation-dominated epoch, but grow
rapidly once the matter-dominated epoch begins, provided their
wavelength is less than the horizon scale at the time $t_{EQ}$ of
matter-radiation equality. The growth of the amplitude scales with
the scale factor of the universe
\begin{equation}
\left(\frac{\delta\rho}{\rho}\right)_R \sim
\left(\frac{R}{R_{EQ}}\right)\
\left(\frac{\delta\rho}{\rho}\right)_0 \label{eq:pp}
\end{equation}
provided the perturbation is small and one remains in the linear
regime. After matter dominates radiation, the Robertson-Walker
scale factor of the universe is given by
\begin{equation}
R(t) = R_0 \left(\mathrm{sinh}\ \frac{3}{2}\, H_\infty
t\right)^{2/3} \ . \label{eq:qq}
\end{equation}
When the sinh factor equals unity, one has equal amounts of
ordinary (dark plus baryonic) matter and dark energy. We denote
this point in time with a subscript $\Lambda$
\begin{equation}
\mathrm{sinh}\ \frac{3}{2}\, H_\infty t_\Lambda = 1
\label{eq:rr}
\end{equation}
and define this as ``cosmological freezeout."  For later times,
when dark energy is dominant, the growth of fluctuations will
cease, and again be frozen in place. Our own universe is in this
state of transition, with the present time $t_0$ given by
\begin{equation}
\mathrm{tanh}\ \frac{3}{2}\, H_\infty t_0 = \sqrt{\Omega_\Lambda}
\approx 0.84 \ . \label{eq:ss}
\end{equation}
It will in general suffice to equate the present time  $t_0$ with
$t_\Lambda$.

The total amount of growth of initial perturbations, from
matter-radiation equality to late times, is therefore simply given
(assuming linearity) by the redshift factor between $t_{EQ}$ and
$t_\Lambda$. Putting in the numbers for our universe, for
short-wavelength modes, one finds
\begin{equation}
\left(\frac{\delta\rho}{\rho}\right)_\Lambda \sim
\left(\frac{R_\Lambda}{R_{EQ}}\right)
\left(\frac{\delta\rho}{\rho}\right)_0 = (1+z)_{EQ}
\left(\frac{\delta\rho}{\rho}\right)_0 \sim
\left(\frac{T_{EQ}}{T_\Lambda}\right)
\left(\frac{\delta\rho}{\rho}\right)_0 \sim (3 \times 10^4)\,
\left(\frac{\delta\rho}{\rho}\right)_0 \ .\label{eq:tt}
\end{equation}
This marginally contradicts the linearity assumption.  Therefore
for fluctuations of wavelength large compared to the critical
wavelength $\lambda_{EQ}$, the total amount of growth will remain
in the linear regime. Consequently we expect that the largest
structures exhibiting very high density contrast will be limited
in size to roughly $\lambda_{EQ}$, defined as the wavelength or
frequency (in comoving conformally flat coordinates) which is
comparable to the horizon scale at $t_{EQ}$. Since
\begin{equation}
\lambda \sim \int^t_0 \frac{dt^\prime}{R(t^\prime)} =
\frac{3t}{R(t)} \sim t^{1/3} \sim R(t)^{1/2} \label{eq:uu}
\end{equation}
it follows that the physical size of this structure at present is
\begin{equation}
r_0 = R_0 \lambda_{EQ} = \left[ \frac{12\,
t_{EQ}}{H^2_\infty}\right]^{1/3} \cong H^{-1}_\infty \left(
\frac{8\, t_{EQ}}{t_\Lambda}\right)^{1/3} = \frac
{8^{1/3}H^{-1}_\infty}{(1+z)^{1/2}_{EQ}} \label{eq:vv}
\end{equation}
which scales as the inverse square root of the redshift factor, as
shown. For our universe, this implies that the density contrast
should be small on scales larger than about 1/400 of the size of
the universe, characterized above by $H^{-1}_\infty$, and large on
scales smaller than that.  This is consistent with what is
observed.

As we mentioned above, we expect that these gross features of this
structure formation are (for the $\Lambda CDM$ scenario)
determined by the cold dark matter which is the dominant component
of the matter density. On the other hand, the structure on smaller
scales may crucially depend upon the baryonic component of the
ordinary matter, because, according to our WIMP hypothesis, the
dark matter component acts as a collisionless dilute gas, while
the baryonic component is more susceptible to nongravitational
dissipative mechanisms.

\begin{figure}[htb]
\centering
\includegraphics{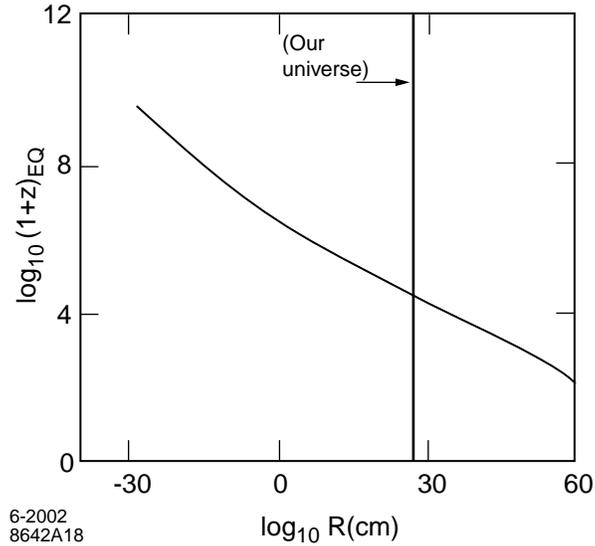}
\caption[*]{\baselineskip=14pt Dependence of the amplification of
primordial fluctuations, $R_\Lambda/R_{EQ} = (1+z)_{EQ}$ upon
$\log_{10}R$. \label{fig:16}}
\end{figure}

We may now investigate how much things change as the radius $R$ of
the universe is varied. We have assumed (quite arbitrarily) that
the primordial fluctuation scale is $\sim 2 \times 10^{-5}$
independent of $R$. As described above, this fluctuation in
general grows linearly with $R$ from the time $t_{EQ}$ of
matter-radiation equality until the time $t_\Lambda$ at which dark
matter and dark energy (cosmological constant) contribute equally
to the Hubble expansion. The former is given in Fig.~\ref{fig:15}
and Eq. (\ref{eq:ll}), while the latter is given by Eq.
(\ref{eq:rr}).
\begin{equation}
T_{EQ} \sim \frac{20\, v^2}{M_{pl}\alpha^n} \qquad \qquad
t_\Lambda \sim H^{-1}_\infty \equiv R \ . \label{eq:ww}
\end{equation}
We must convert $T_{EQ}$ to $t_{EQ}$ using the fundamental
relationship between them, valid in the radiation-dominated epoch:
\begin{equation}
t_{EQ} \sim \frac{M_{pl}}{T^2_{EQ}} \ . \label{eq:xx}
\end{equation}
We thereby obtain
\begin{equation}
\left(\frac{\delta\rho}{\rho}\right)_\Lambda \sim
\left(\frac{t_\Lambda}{t_{EQ}}\right)^{2/3}
\left(\frac{\delta\rho}{\rho}\right)_0 \sim \left(\frac{R\,
T^2_{EQ}}{M_{pl}}\right)^{2/3}\left(\frac{\delta\rho}{\rho}\right)_0
\equiv (1+z)_{EQ}\left(\frac{\delta\rho}{\rho_0}\right) \ .
\label{eq:yy}
\end{equation}
The size $r_0$ of the largest structures, relative to the size $R$
of the universe, follows from Eq. (\ref{eq:vv}) and has a similar
form:
\begin{equation}
r_0 \sim \left(\frac{t_{EQ}}{t_\Lambda}\right)^{1/3} R \sim
\left(\frac{R\, T^2_{EQ}}{M_{pl}}\right)^{1/3} R \equiv
(1+z)^{-1/2}_{EQ} R  . \label{eq:zz}
\end{equation}
The $R$ dependence of the red-shift factor $(1+z)_{EQ}$  is shown
in Fig.~\ref{fig:16}. We see that for a rather large bandwidth, of
order 10 to 15 powers of ten, the amplification of the primordial
perturbations is within an order of magnitude of what is present
in our own universe. We also recall that for universes larger than
$10^{-4}$ of ours, the baryonic fraction $\Omega_B/\Omega_X$ is
small, but not negligibly so. Therefore we may surmise that,
whatever the mechanism is that creates black holes in the centers
of galaxies, it will probably still be operative in this class of
universes as well.  However, the basis for this conclusion is very
fragile, since it rests upon our assumptions of the properties of
dark matter.

\section{Emergent Cosmology}\label{sec:5}

If the ensemble of universes we have been considering actually
exists, then there are anthropic consequences, as mentioned in the
introduction. From the behavior found in the previous section, we
may conclude that if the number of universes per octave (factor
two) in radius $R$ is large compared to unity, then it is
reasonable that we should be present in the ensemble
\cite{Muller}. There is a caveat; if the mean number of planets
per universe (of our size $R$) which are appropriate for the
support of life as we know it is small compared to unity, then the
number of universes per octave needed to make reasonable our
existence must be correspondingly increased. The planetary
situation is not well understood \cite{ref:BW}, so this option is
not academic. But either way, it would not seem outrageous that
enough universes exist to take care of the problem.

There are probably as many models of multiverses as there are
practitioners foolish enough to deal with the idea. In this
section we shall play with a specific model, motivated by the idea
of emergent field theories, a concept born from analogies with
condensed matter physics \cite{ref:h}. The model will also be
related to the ideas of evolutionary cosmology developed by Smolin
\cite{ref:e}, albeit in a more deterministic framework. As
mentioned in the introduction, our reason for indulging in this
fantasy is to try to obtain some guidance in the search for a
satisfactory microscopic emergent theory.

The basic premise underlying the emergence approach is that the
vacuum of particle physics and cosmology is analogous to a quantum
liquid in equilibrium at very low temperature. Such a system has
essentially zero pressure. But the measure of vacuum pressure is
the cosmological constant itself, explaining not only why it
should be zero, but why it is not quite zero: a droplet of vacuum
of finite size will have pressure due to surface effects. This is
just what happens in the DeSitter universe (\cf\
Eq.~(\ref{eq:bb})).

Chapline {\em et al.} \cite{ref:g}, and Mazur and Mottola
\cite{ref:f}, have recently carried this notion further, and argue
that a black hole is to be considered a droplet of quantum liquid,
with a nonsingular interior which is in fact static DeSitter
space. The value of the cosmological constant in the interior
differs from its value exterior to the horizon and serves as a
kind of order parameter. In what we have described, this is
generalized to all the standard model parameters, which evidently
are also discontinuous across the horizon. This picture is
ready-made for the cosmological setting in which we find
ourselves: not only does our universe contain a large number of
``daughter" black-hole fluid droplets, but our universe itself can
be considered the interior of a much bigger droplet, which
presumably exists, along with many other ``sister" droplets, in a
much larger ``mother" universe. From this starting point, one
easily sees that a genealogy can be defined. The properties of
mother and daughter universes will depend upon how different in
size they are from our own, and how differently the physics works
at those size scales. It is this question that we take up in this
section, building upon what was learned in the previous sections.,
We shall not venture very far beyond one generation in either
direction; there will be more than enough uncertainty at this
level.

It is easiest and most direct to first consider the daughter
universes, because there are some data. There are at least two
kinds of daughters---the supermassive, galactic black holes, of
horizon size roughly 15 orders of magnitude smaller than the size
of our universe, and the stellar-size black holes which are six to
eight orders of magnitude smaller still. Despite the greater
uncertainty in the underlying astrophysics, we specialize to the
former because they are closest to us in size.

Rather than characterizing the black hole size by its horizon
radius, it is also useful to give it in terms of the volume of
comoving matter needed to form the black hole, which we assume is
baryonic in origin. In our universe the mass of such black holes
is in the range of $10^6$ to $10^9$ solar masses , or $10^{53}$ to
$10^{56}$ proton masses, out of a total of about $10^{79}$ in the
universe. So the fraction by volume of total comoving baryonic
matter that ends up in one of these black holes is $10^{-23}$ to
$10^{-26}$. Taking a cube root gives the fraction in linear scale
of roughly $10^{-8}$ to $10^{-9}$. This should be compared with
the fraction in linear scale of about $10^{-3}$ for the largest
scale structures found in our universe.

Our main purpose in spinning out these numbers is to try to infer
the most likely size of our mother universe. Evidently the first
rough guess would be 15 powers of ten larger than our own. But by
the time one goes out those fifteen orders of magnitude, the
cosmology has significantly changed, and it is possible that one
must go even further. Let us review what was learned in the
previous section for the cosmology of a candidate mother universe,
say, of radius $10^{45}$ cm.

\begin{figure}[htb]
\centering
\includegraphics{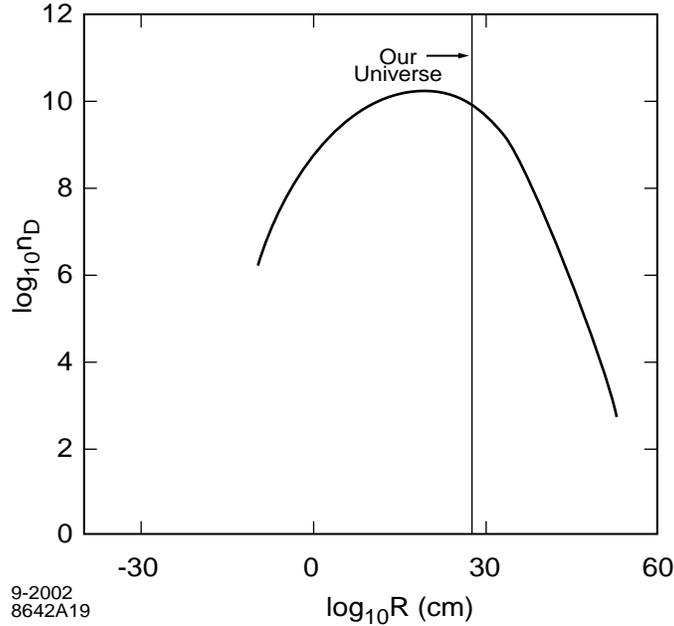}
\caption[*]{\baselineskip=14pt A guess for the dependence of
fertility, {\em i.e.} the number $n_D$ of daughter universes per
mother, upon size $R$. \label{fig:17}}
\end{figure}

The early evolution of a mother universe of this size would be
similar to our universe. Nucleosynthesis would occur in one of
three possible scenarios (\cf\  Fig.~\ref{fig:13}), but in all
cases hydrogen would predominate in the long run. The three cases
are distinguished by the nonexistence of deuterium and/or the
relative abundance of primordial helium. As the temperature
decreased, decoupling of matter from radiation would occur during
the radiation-dominated epoch (\cf\  Fig.~\ref{fig:15}). This
means that the growth of density contrast would from the outset
involve both the baryonic matter and the dark matter. However, the
growth factor, which scales with the redshift at the time $t_{EQ}$
of matter-radiation equality, is less by about a factor ten than
for our universe (\cf\  Fig.~\ref{fig:16}). In addition, the ratio
of baryonic to dark matter is much less; instead of ten percent,
the number is somewhere between 3 percent and 0.01 percent. All of
these features will make baryonic structure formation more
difficult. What is most important for our consideration here is
whether these mother universes can give birth to daughter black
holes. Baryonic matter has to aggregate in the potential wells
created by the dark matter and undergo gravitational collapse.
While it apparently is more difficult for this to happen, it is
not clear that the number of black holes that might be created is
in fact small compared to unity, when for our universe the
corresponding number (for ``galactic" black holes) is $10^{10}$ or
so. We shall make a guess that the fertility curve looks something
like what is depicted in Fig.~\ref{fig:17}; this would allow
mother universes to be present, but make it unlikely that
grandmothers exist. But we must emphasize the many huge
uncertainties involved, not the least of which is the assumption
that the primordial fluctuation spectrum does not depend upon $R$.

\begin{figure}[htb]
\centering
\includegraphics{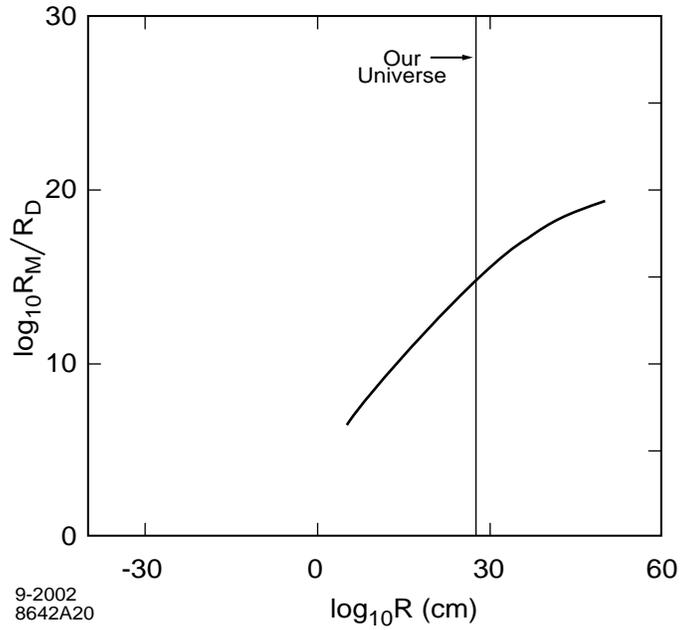}
\caption[*]{\baselineskip=14pt A guess for the ratio of the
average size $R_M$ of a mother universe, assumed to be a
supermassive black hole interior, to the size $R$ of its daughter,
versus $R$. \label{fig:18}}
\end{figure}

Relative to our universe, the fraction of matter in the mother
universe which is baryonic is, as already mentioned, less than for
our universe. This might affect not only the frequency of
occurrence of black-hole formation, but also the size
distribution. We do not try to estimate the effect, mainly out of
lack of competence. But it is likely that the ratio of size of
mother to daughter indeed grows with overall scale $R$, in the way
sketched out in Fig.~\ref{fig:18}. But we emphasize that we are
approaching a level of almost complete guesswork.

Despite all these uncertainties, it seems relatively safe to
conclude, given our assumptions, that the model of nested black
holes for the multiverse allows at most one or two generations of
parents, with a number of sister universes small compared to
$10^{10}$, the number of (galactic) black hole daughters in our
universe. It seems very unreasonable to assume a large number of
``ancestor" generations, unless the primordial density
fluctuations were to increase is magnitude with $R$. However,
intuitively we would if anything expect the opposite to occur.

Since the size distribution of sister universes span a few factors
ten, the fraction with size close enough to our universe to in
principle support life as we know it will be a few powers of ten
less than the total population of sisters. This means that the
total number of universes in such a nested-black-hole multiverse
which could support life as we know it is bounded above by ten to
a small power. It follows that the overall number of planets in
the multiverse that are candidates for habitable environments is
not all that different (on a logarithmic scale) from the number in
our own universe.

The above inferences are rather strong, and therefore invite an
additional critical look: are these conclusions avoidable? In such
a soft topic as the contents of this paper, the answer is almost
certainly yes. One assumption we have been making, mainly from a
desire for simplicity and definiteness, is that the ensemble of
universes we consider is parametrized only by the size $R$ and
nothing else. The other cosmological parameters, such as magnitude
of the baryon asymmetry and/or the magnitude of the primordial
density fluctuations, may well represent independent initial
conditions, unconstrained or at best loosely constrained by the
value of the size parameter $R$. In such cases the conclusions
about abundances of mother and grandmother universes are
inoperative. Exploration of such alternatives seems however to be
premature, and in any case beyond the scope of this paper.

\section{Lessons and Challenges} \label{sec:6}

While everything we have discussed is very speculative, it must be
admitted that, given the starting hypothesis of size-dependent
standard-model parameters, we have been able to look at old
questions from a somewhat different perspective. This in itself
can be a benefit, inasmuch as a fresh point of view is often a key
to making progress. And in fact there are some of the classic Big
Questions for which partial answers can be set forward:

\medskip
\noindent 1. {\em Why are there such hierarchies in scale amongst
standard model parameters?}

This question includes the classic ``hierarchy problem", namely
the smallness of the electroweak scale $v$ relative to the
Planck/GUT scale, which stimulates the introduction of weak-scale
supersymmetry by so many practitioners. It also includes the
question of why the electron mass is so much smaller than the
top-quark mass. And it even includes the question of why the
cosmological-constant scale is so much smaller than the QCD and
electroweak scales, not to mention the Planck/GUT scale.

The answer to this general question, given the multiverse
hypothesis, may be that for most universes in the ensemble there
is no such huge hierarchy. If the size distribution of universes
is maximum for relatively small values of $R$, say the GUT scale
or smaller, then the typical universe has no large disparity of
scales. Only the large, rare, universes like ours enjoy that
property as a consequence of the assumed scaling behavior of
parameters (which of course must eventually be explained).

\medskip
\noindent 2. {\em Why is the fine-structure constant 1/137 so
small?}

The answer to this famous old question is the same as above: in
small universes $\alpha$ is not small; only in large ones like
ours is it small. There are corollaries which are answered in the
same way. The most immediate is the more modern version of the
above question: why are the gauge coupling constants at the GUT
unification scale so small? And directly related to this question
is why the QCD scale $\Lambda_{QCD}$ is so small relative to the
GUT scale. All these questions are answered in the same way:
because we live in a very large universe.

\medskip
\noindent 3. {\em Why is our universe so large?}

This is the obvious follow-up question to the previous ones. And
the answer to this is weakly anthropic: our universe is large
because we inhabit it. The discussions in the previous sections
show it could not be otherwise, given the scaling assumptions
underlying this note.

But in addition to these questions, there is the most important
one, which remains without much of an answer:

\medskip
\noindent 4. {\em Why should the assumed ``fixed-point" scaling
behavior be true?}

One response is that it (Fig.~\ref{fig:3}) looks just as credible
as the conventional-wisdom alternative (Fig.~\ref{fig:1})---which
in itself sheds no light on the above questions. But at best this
response is highly subjective and leaves much to be desired. To
give a more satisfactory reply would be to relate the scaling
behavior to the microscopic theory. This has not been done. But
there are some interesting guidelines which the assumed behavior
suggests. One concerns the limit of the standard model for
infinite $R$. In that limit all dimensionless coupling constants
vanish, and the standard model becomes trivial \cite{ref:b}. In
other words, the presence of nontrivial interactions of the
particles with each other depends upon the existence of a
nonvanishing cosmological constant. In the emergent, ``gravastar"
scenario, this states that the standard-model interactions are
present only because of the presence of the DeSitter horizon, in
the neighborhood of which exists new, beyond-the-standard-model
physics. It is as if all the standard-model forces are in some
sense Casimir effects. However, the standard kind of Casimir
effect, which depends upon size of the system as an inverse power,
will not do the trick. There are terms in the standard-model
action, such as the Higgs mass term and the cosmological-constant
term itself, which do have the typical behavior. But most of them,
after appropriate rescaling of fields, depend only logarithmically
upon the size parameter $R$. To see this, write schematically the
standard model Lagrangian as
\begin{equation}
\L = F^2 + \bar \psi \Dslash \psi + (D\phi)^2 + g\bar \psi\psi\phi
+ g^2\phi^4 - \mu M\phi^2 \label{eq:qa}
\end{equation}
where the first three terms are gauge, fermion, and Higgs kinetic
energy terms, and the last terms are Yukawa, quartic Higgs, and
Higgs mass terms respectively.\footnote{We here conjecture, as in
our previous note \cite{ref:b}, that the Higgs mass is the
geometric mean of the cosmological and Planck/GUT scales.}   The
covariant derivative is
\begin{equation}
D = \partial - g A \label{eq:qb}
\end{equation}
and $g$ is a generic label for gauge or Higgs coupling; we take
$\lambda \sim g^2$ because the assumed $R$ dependence is then
universal.  Under the rescalings
\begin{equation}
A = g^{-1}\!\!\!\! \lowsim A , \quad \psi = g^{-1}
\!\!\!\lowsim\psi , \quad \phi = g^{-1}\!\!\! \lowsim\phi , \quad
\lowsim D = \partial\  - \!\!\!\lowsim A \label{eq:qc}
\end{equation}
we find
\begin{eqnarray}
\L &=& g^{-2} \left\{ \lowsim F\!\!^2 +
\lowsim{\bar\psi}\!\!\!\lowsim D\!\!\!\lowsim \psi +
(\!\!\!\lowsim D\!\!\!\lowsim\phi)^2 +
\lowsim{\bar\psi}\!\!\!\lowsim\psi\!\!\!\lowsim\phi +
\lowsim\phi^4 - \mu M\!\!\!\lowsim \phi^2\right\}\nonumber \\ & &
\sim (\log\, MR)\lowsim\L \ . \label{eq:qd}
\end{eqnarray}
The action is
\begin{equation}
S = \frac{1}{\hbar} \int d^4x\ \L =  \frac{1}{\hbar(R)} \int d^4x
\lowsim\L \label{eq:qe}
\end{equation}
with
\begin{equation}
\hbar(R) \sim \frac{1}{(\ell n\, MR)} . \label{eq:qf}
\end{equation}
 The entire Lagrangian density gets multiplied by a factor
$\ell n\, M_{pl} R$, as if the Planck constant itself is
scale-dependent, vanishing in the limit of infinite $R$.

We have not considered in this note such a possibility, and have
in fact essentially set the Planck constant, the speed of light,
and the Planck mass to unity, not allowing them to vary with $R$.
As long as the universes in the multiverse are causally
disconnected from each other, this can be defended as no more than
a convention in the choice of units \cite{Duff}. However, if there
is a connection between the universes, such as in the
nested-black-hole scenario, then it is no longer obvious that this
is a safe assumption. Relaxation of such an assumption might in
fact lead to additional insight. However, exploration of this
possibility is beyond the scope of this work.

In addition to the questions above, for which our approach might
provide some insight, there are others for which our present lack
of understanding is highlighted, and which need better answers in
order to sharpen the consequences of the scaling assumptions which
we have made. These include:

\medskip
\noindent 5.  {\em What is the mechanism by which the electron and
the light quarks get their mass?}

This is often viewed as a minor detail in the grand scheme of
standard model problems. But in the context of this note, the lack
of understanding of the origin of light quark and lepton masses is
translated into relatively great uncertainty in the understanding
of the relationship of our universe to other universes of
different size.

\medskip
\noindent 6.  {\em What is the nature of the dark matter?}

\noindent 7. {\em What is the origin of the baryon asymmetry, and
what determines its magnitude of $\eta = 3 \times 10^{-10}$?}

\medskip
\noindent 8. {\em Why is the value of the primordial density
fluctuations $(\delta \rho/\rho)_0$ equal to $2 \times 10^{-5}$?}

These three cosmological questions are hardly novel \cite{Rees6};
they are evidently crucial to better understanding the properties
of the ensemble of universes we consider. Perhaps the only novelty
is that we omit (here) the question of the ``small" cosmological
constant, usually added to the above list.

It is also worth noting that, while the flatness problem (why
$\Omega = 1$) and the scale-invariant spectrum of primordial
fluctuations represent something of a triumph for the idea of
inflation, there remains no good answer to the eighth question:
the {\em  magnitude} of the primordial fluctuation spectrum is
simply fit to the data, and not understood at all from more
fundamental considerations. And, as discussed at the end of the
previous section, it is possible that these parameters should be
considered as independent characterizations of members of the
ensemble of universes, {\em i.e.} as initial conditions not
strongly dependent upon the size parameter $R$.

Finally, there are the lessons, if any, which are learned from
this exercise that may be applied to the hypothesis of emergence.
The idea of emergence provides some motivation for the scaling
behavior assumed from the beginning of this note. But it has many
daunting problems associated with it:

\medskip
\noindent 1. {\em Why are violations of Lorentz covariance so
small?}

Condensed-matter analogs of emergence suggest in general that
symmetries such as Lorentz covariance are just low energy
approximations. At high enough energies deviations are to be
expected. But experiment severely limits such deviations. For
example, noncovariant corrections to charge-renormalization, an
ultraviolet-sensitive quantity, are limited \cite{QED} to less
than one part in $10^{31}$. This comprises a staggeringly
restrictive constraint. It would seem essential that there be a
very small parameter which characterizes the violations. And the
scaling behavior of parameters studied here suggests that a
necessary (but far from sufficient) condition for the
Lorentz-violating terms in the Lagrangian is that they scale as
inverse powers of the radius $R$ of the universe. If this is so,
then very small universes exhibit very little symmetry, while the
very large ones like our own exhibit Lorentz symmetry, {\em etc.},
with very small corrections.

\medskip
\noindent 2.  {\em What is the structure of event horizons?}

In the nested-black-hole, or ``gravastar" scenario, there is ``new
physics" at horizons. This is endemic in the condensed-matter
analogues \cite{ref:i}. And in our picture, standard-model
parameters (including the cosmological constant) are discontinuous
across horizons, indicating that at the surface of discontinuity
conventional-physics descriptions of what is going on are
incomplete. There also appear to be violations of the weak energy
conditions of classical general relativity \cite{Israel}. One
manifestation of this appears to be that there are large classes
of null geodesics (in particular those which have nontrivial
transverse motion) which are ``bound" to the horizon. There is a
nontrivial problem here of providing a consistent description.

In addition, if our universe is to be regarded as the DeSitter
interior of a gravastar, then there must be in our universe
preferred comoving observers, presumably not ourselves, with
respect to which there is the ``physical" horizon associated with
our black-hole interior. It then becomes an interesting question
as to where we should regard ourselves relative to these central
observers: how far away are they, and in what direction? Might
there be observational issues associated with this preferred
center of our universe? While these are quite interesting
questions, they also lie beyond the scope of this note.

\medskip
\noindent 3.  {\em How are gravastars formed?}

If the gravastar picture is in fact viable, then there must be a
time evolution of the ``new physics" which is associated with the
horizon. But for large black holes, it is hard to find an
intrinsic, local parameter associated with the horizon, because
classically it can be regarded as an artifact associated with a
choice of coordinate systems. In the emergence scenario, general
covariance is only a low energy approximation. This implies that
the description of gravastar formation will require a ``best"
choice of coordinates. What should be chosen?

This is only one of the difficult issues involving gravastar
formation. Another involves rotation: no ``eternal rotating
gravastar" generalization of the nonrotating case has been
found.\footnote{\baselineskip=14pt A default option is to invoke
an ``eye of the hurricane" model. Choose the Kerr metric for the
exterior, and static DeSitter space for the interior. Then build
an appropriate interpolating boundary layer with an exotic
spacetime, which probably contains vorticity.}  In nested-black
hole cosmologies, it is necessary that all the black-hole
universes (including ours) are characterized by a value of spin as
well as mass. This is not only a complication, but also an
opportunity for linking standard model discrete symmetry
violations, in particular $CP$, to the existence of a spin axis
for the new physics at the DeSitter horizon---physics which
presumably controls the nontrivial interaction features of the
standard model.

On the more positive side, some insight on the history of
gravastar formation might be gleaned by comparing the formation of
a daughter black hole with the formation of our own parent
universe. The characteristic time for the formation of the
daughter can be easily taken to be at the very least many millions
of years, a timescale much larger than the size of the gravastar.
If we assume the same for our parent universe, it follows that the
formation time for our universe should be considered to be much
larger than the size parameter $R$---in other words orders of
magnitude larger than $10^{10}$ years \cite{Lenny}. This might
imply that the formation time of the ``new physics" on the
DeSitter horizon likewise is long compared to $10^{10}$ years.
Some kind of cosmological ``bounce" scenario \cite{BD} might have
the best chance of providing a concrete implementation of this
inference.

\medskip
\noindent 4.  {\em What is the microscopic physics underlying the
emergence scenario?}

This question remains unanswered. Necessary conditions are that
the gauge bosons of the standard model, as well as the graviton,
should be considered collective modes of the presumed ``quantum
liquid" vacuum. They quite likely should be all considered
Goldstone modes \cite{ref:b,UCLA} associated with various kinds of
spontaneous symmetry breakdown. The pattern of internal
symmetries, especially in the fermion representations, must be an
essential clue.

Finding the answer to this last question may well afford the best
chance of turning the very speculative material in this note into
something considerably more concrete.

\section*{Acknowledgments}

It is a pleasure to thank my colleagues at Stanford for many
helpful discussions, especially R. Adler, P. Chen, and M.
Weinstein.  I also thank S. Beane and M. Savage for valuable
discussions on the chiral limit of nuclear physics.

\end{document}